\documentclass[sigconf]{acmart}
\renewcommand\footnotetextcopyrightpermission[1]{} %
\usepackage[utf8]{inputenc}
\usepackage{subcaption}%
\usepackage{siunitx}%
\usepackage{bm}
\usepackage[acronym]{glossaries}%
\newacronym{cpg}{CPG}{central pattern generator}

\setcopyright{acmcopyright}

\acmConference[NICE '20]{Neuro-inspired Computational Elements Workshop}{March 17--20, 2020}{Heidelberg, Germany}

\begin{document}

\title{Adaptive control for hindlimb locomotion in a simulated mouse through temporal cerebellar learning} %

\author{T. P. Jensen}
\email{thompa@elektro.dtu.dk}
\affiliation{%
    \institution{Automation and Control, Department of Electrical Engineering, Technical University of Denmark}
    \streetaddress{Elektrovej, Building 326}
    \city{Kgs. Lyngby}
    \postcode{2800}
    \country{Denmark}
}

\author{S. Tata}
\email{shravan.tataramalingasetty@epfl.ch}
\affiliation{%
    \institution{Biorobotics Laboratory, School of Engineering, École Polytechnique Fédérale de Lausanne}
    \streetaddress{Station 14}
    \city{Lausanne}
    \postcode{1015}
    \country{Switzerland}
}

\author{A. J. Ijspeert}
\email{auke.ijspeert@epfl.ch}
\affiliation{%
    \institution{Biorobotics Laboratory, School of Engineering, École Polytechnique Fédérale de Lausanne}
    \streetaddress{Station 14}
    \city{Lausanne}
    \postcode{1015}
    \country{Switzerland}
}

\author{S. Tolu}
\email{stolu@elektro.dtu.dk}
\affiliation{%
    \institution{Automation and Control, Department of Electrical Engineering, Technical University of Denmark}
    \streetaddress{Elektrovej, Building 326}
    \city{Kgs. Lyngby}
    \postcode{2800}
    \country{Denmark}
}

\begin{abstract} %
Human beings and other vertebrates show remarkable performance and efficiency in locomotion, but the functioning of their biological control systems for locomotion is still only partially understood. The basic patterns and timing for locomotion are provided by a central pattern generator (CPG) in the spinal cord. The cerebellum is known to play an important role in adaptive locomotion. Recent studies have given insights into the error signals responsible for driving the cerebellar adaptation in locomotion. However, the question of how the cerebellar output influences the gait remains unanswered. We hypothesize that the cerebellar correction is applied to the pattern formation part of the CPG.
Here, a bio-inspired control system for adaptive locomotion of the musculoskeletal system of the mouse is presented, where a cerebellar-like module adapts the step time by using the double support interlimb asymmetry as a temporal teaching signal. The control system is tested on a simulated mouse in a split-belt treadmill setup similar to those used in experiments with real mice.
The results show adaptive locomotion behavior in the interlimb parameters similar to that seen in humans and mice. The control system adaptively decreases the double support asymmetry that occurs due to environmental perturbations in the split-belt protocol.
\end{abstract}

\keywords{Bio control, learning algorithms, adaptive locomotion, brain models, motor control}

\maketitle

\section{Introduction}

The locomotion of vertebrates has been studied for a very long time in research areas of biology and neurology, and it has also attracted interest from roboticists because of the inherent ability of biological control systems to adapt and being robust to disturbances within the environment. %
The consensus in the neurological literature is that the spinal cord contains a \gls{cpg} responsible for generating the timing and patterns of muscle activation signals for locomotion \cite{Dietz2003,Kiehn2006,Takakusaki2013}. 
Examples of robotic control systems with \gls{cpg}s include quadrupedal robots \cite{Espinal2016a}, robot snakes \cite{wu2010cpg}, salamanders \cite{Ijspeert2007} and bipedal robots \cite{Matsubara2005,Fujiki2013}.

Within the brain, the cerebellum has a key role for adaptive locomotion \cite{Morton2004}. While the \gls{cpg} provides the basic control patterns to generate the gait, the cerebellum is required to adapt the patterns in case of changes or perturbations in the environment \cite{Morton2006}. Evidence suggests \cite{Malone2012, Mawase2013} that the cerebellum works as an adaptive feed-forward controller for the locomotion task since an after-effect is present when the perturbation is removed. The fundamental research question is how adaptive locomotion is achieved in vertebrates, and whether we can apply it to robots. In this work, we aim to identify the error signal responsible for driving the adaptation, and to locate where the correction is given by the cerebellum. It is debated whether the cerebellum gives corrections in terms of a reference trajectory for the limbs or rather with some higher objectives in mind. %
Considering that the spinal \gls{cpg} is a complex network of interneurons, with groups that are responsible for rhythm generation, pattern generation, and motor output respectively \cite{Takakusaki2013}, it is natural to speculate to which group the cerebellar correction applies. In this paper, we hypothesize that for rodents and humans in locomotion, the double support asymmetry is used as an error signal in the cerebellum to give corrections to the pattern formation part of the \gls{cpg}.

Some researchers performed trials with humans on a split-belt that consist of a treadmill with two belts that are individually controlled \cite{Reisman2005, Morton2006, Malone2012}, to analyze the adaptive behavior. Morton describe the adaptation of gait parameters \cite{Morton2006}. The results showed two types of adaptation, described as reactive and predictive. The reactive adaptation happened rapidly and returned promptly to the baseline values in the post-adaptation period. The predictive feed-forward adaptation occurred more slowly than the reactive one and it had an after-effect in the post-adaptation period. Their findings for patients with cerebellar damage showed no loss of the reactive adaptation. The predictive adaptation was, however, much less (or absent) compared to healthy patients. They identified the reactive adaptation as impacting the intralimb parameters, stance time, and length. The predictive adaptation was seen in interlimb parameters, step length, and double support time.

In a series of articles, it has been proposed that the locomotor cerebellar adaptation in humans consists of a spatial and a temporal component. Experiments during split-belt trials show no impact on a temporal adaptation, when the subjects are distracted \cite{Malone2010}. The spatial adaptation, however, suffers from the distraction, which leads to the conclusion that two different neural circuits are involved in the learning process to reduce the step length asymmetry. These were shown to adapt at different rates and in an experiment it was possible to circumvent the adaptation of the temporal component \cite{Malone2012}. For the temporal component, the authors proposed that the double support asymmetry is utilized as a temporal error signal for adaptation in human locomotion.
It has been shown that cerebellar adaptations can be separated into spatial and temporal contributions also in mice \cite{Darmohray2019}. Their results suggest that the adaptation is primarily on the front limbs and that the hindlimbs might simply adjust to match the patterns of the front limbs. It is argued that the cerebellum adapts to reduce the step length asymmetry, which they decompose into adaptations in spatial and temporal domains. The spatial adaptation affects the asymmetry in the center of oscillation and the temporal adaptation affects the double support asymmetry. Neither of those asymmetries is reduced to zero in their experimental results, rather they are both reduced to give almost no step length asymmetry. The cerebellar adaptations of humans and mice carry many similarities, according to \cite{Darmohray2019}. This hints that the structure of the involved cerebellar circuits might be very similar for rodents and humans, even considering the differences between bipedal and quadrupedal locomotion.

To study the biological control systems for locomotion, Fujiki et al. presented a \gls{cpg} model for controlling the hindlimbs in a simulated rat \cite{Fujiki2018}. The model used a hip reflex for phase resetting in the neural oscillators to achieve reactive adaptation in the split-belt protocol. Cerebellar models are less common in locomotion control. In an earlier study by Fujiki \cite{Fujiki2015}, a bipedal robot is controlled by means of a \gls{cpg} with a similar oscillator model, and a cerebellar-like model to adapt the phase resetting with a foot contact sensor. The cerebellar model used a simple equation for learning based on a phase prediction error. This is the only previous example of a cerebellar-like model for adaptive locomotion in a robot, where the cerebellar correction is not applied to the motor output. For example, in \cite{Massi2019} a cerebellar-like controller was used to give joint torque corrections for locomotion of a quadrupedal robot, by utilizing a reference limb trajectory to generate the error signals.

In this work, we sought to model the cerebellar temporal adaptation in mice locomotion. To do so, a control architecture was designed by including a modified version of the \gls{cpg} by Fujiki et al. \cite{Fujiki2018}. A cerebellar learning rule similar to the one described in \cite{Fujiki2015} was applied, however, the error signal was based on the theories of Malone et al. \cite{Malone2012}. In contrast to \cite{Fujiki2015}, the cerebellar output is applied as a correction signal to the pattern formation part of the \gls{cpg}. To investigate how adaptive locomotion is achieved for the mouse, the musculoskeletal system was simulated using muscle models and controlled by the developed bio-inspired control architecture. 

The organization of the paper is as follows: an overview of the mouse hindlimb simulation is given in section \ref{sec:method} together with a definition of the experimental protocol and a presentation of the bio-inspired control system; in section \ref{sec:results} the results are presented; finally, in section \ref{sec:discussion}, the results are discussed and compared with the ones from real mice and humans to highlight the effect of the selected error signal on the adaptive locomotion behavior.

\section{Methods}\label{sec:method}
\subsection{Experimental setup}
\begin{figure}[h]
    \centering
    \includegraphics[width=0.7\linewidth]{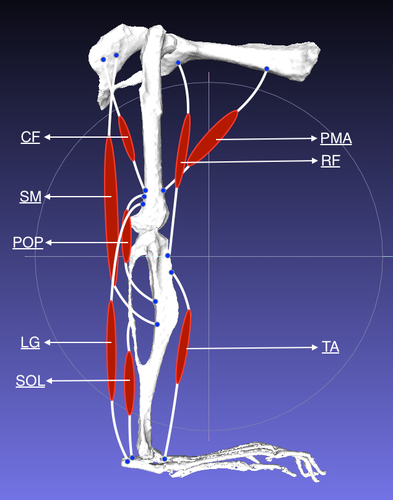}
    \caption{Illustration of the mouse hindlimb model, not showing the exact attachments of the muscles. The muscles that have been modeled was selected from the full set of muscles due to their relevance in forward locomotion. For a more accurate picture see \cite{charles2016muscle}. PMA: psoas major, CF: caudofemoralis, SM: semimembranosus, POP: popliteus, RF: rectus femoris, LG: lateral gastrocnemius, SOL: soleus, TA: tibialis anterior.} %
    \label{fig:hind_limb_muscles_locomotion}
\end{figure}

The experimental setup consists of a simulation of a mouse on a split-belt treadmill in the Webots simulation environment \cite{Michel2004}. Only hindlimb locomotion is considered. The hip, knee and ankle joints of each hindlimb are controlled by a total of 8 muscles consisting of both monoarticular (spanning across one joint) and biarticular (spanning across two joints) muscles. The muscles that are included in the mouse model are shown in Figure \ref{fig:hind_limb_muscles_locomotion}.

The mouse model was based on a high-resolution 3-dimensional scan of a mouse skeleton and the muscles are modeled using Geyer muscle equations \cite{Geyer2003,Geyer2010} with muscle parameters from \cite{charles2016muscle}.

\begin{figure}[tb]
\centering
\includegraphics[width=0.9\linewidth]{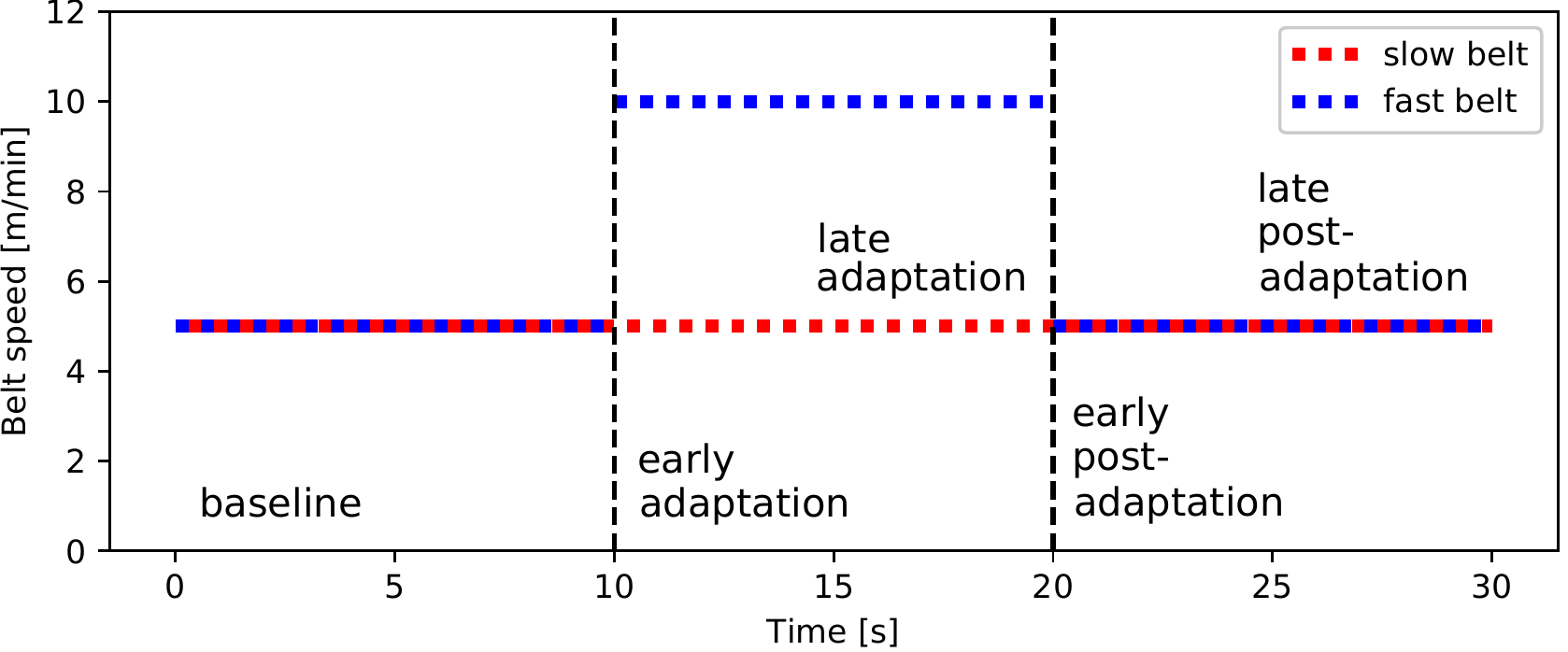}
\caption{The split-belt protocol (2.0x speed ratio), showing the velocity of each belt for one experiment. }
\label{fig:split_belt_protocol}
\end{figure}

The experimental split-belt protocol has been used extensively on real mice, cats, and humans in the literature \cite{Fujiki2018,DAngelo2014,Helm2015}. The setup consists of two separate treadmills, one for each side of the body. By changing the velocity of one of the belts, the gait is perturbed. An example of an experimental trial is illustrated in Figure \ref{fig:split_belt_protocol}. Typically, it consists of a baseline period, where the belt velocities are tied, a split period where the velocity of one of the belts is increased, and finally a period of tied belt velocities. Here, the base belt velocity is selected to \SI{6}{m/min} and speeds of 9, 10.2 and \SI{12}{m/min} are used for the fast belt (1.5x, 1.7x and 2.0x speed ratio). This base velocity was selected by matching belt speed to the \gls{cpg} frequency and the step length in a single belt experiment. To describe the adaptation, the split period is divided into early and late adaptation periods, with the second tied-belt period being divided similarly.

Because the mouse is only walking on the hindlimbs, the spine is fixed in space by a physics plugin in the simulator. This setup is similar to that in \cite{Fujiki2018}, where the rats wore a harness, and the front limbs were supported by a bar.

\begin{figure}[tb]
    \centering
    \includegraphics[width=0.7\linewidth]{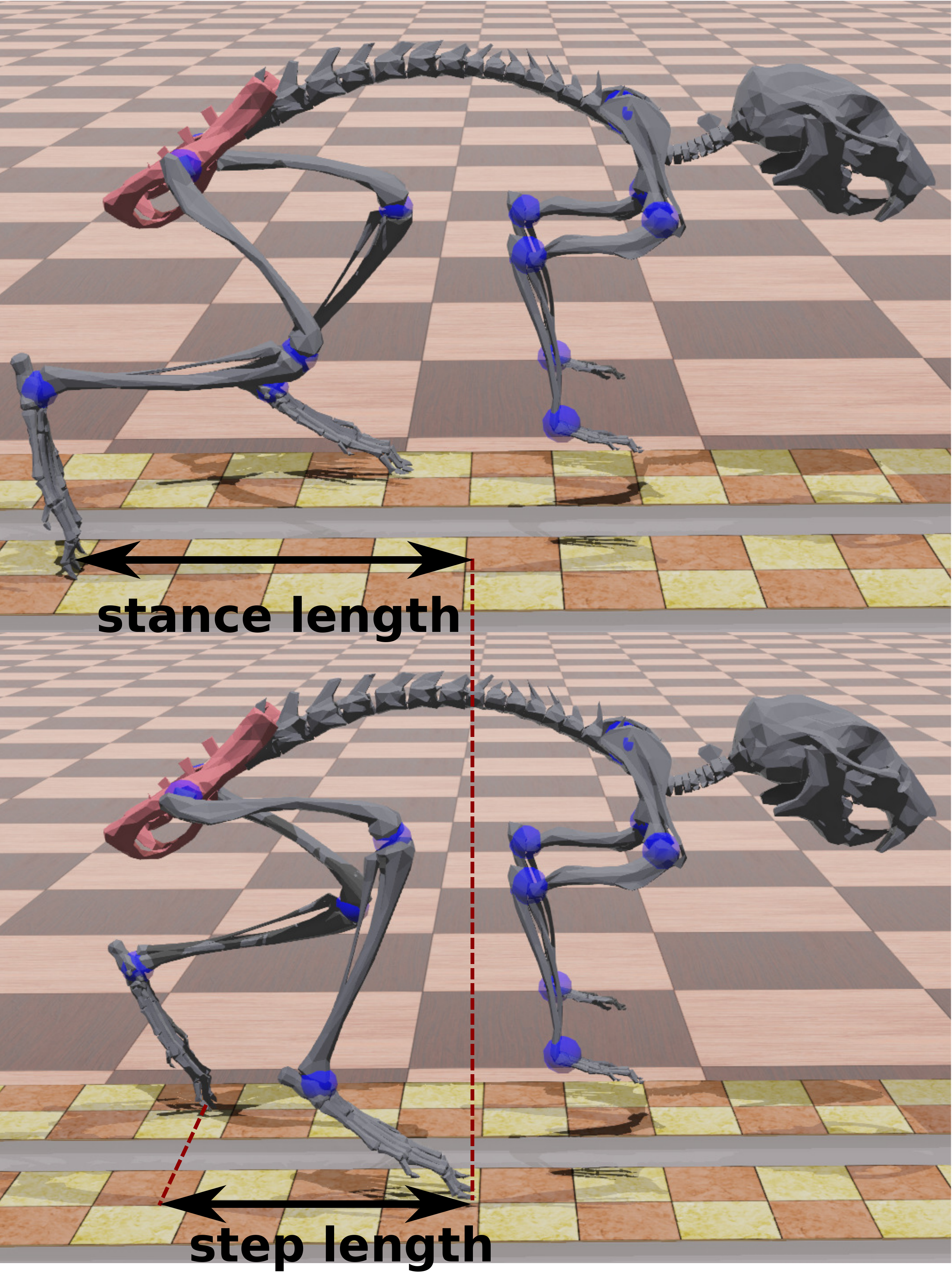}
    \caption{Illustration of the spatial gait parameters, stance length, and step length. Blue dots denote the joints.}
    \label{fig:gait_spatial_parameters}
\end{figure}

\begin{figure}[tb]
    \centering
    \includegraphics[width=\linewidth]{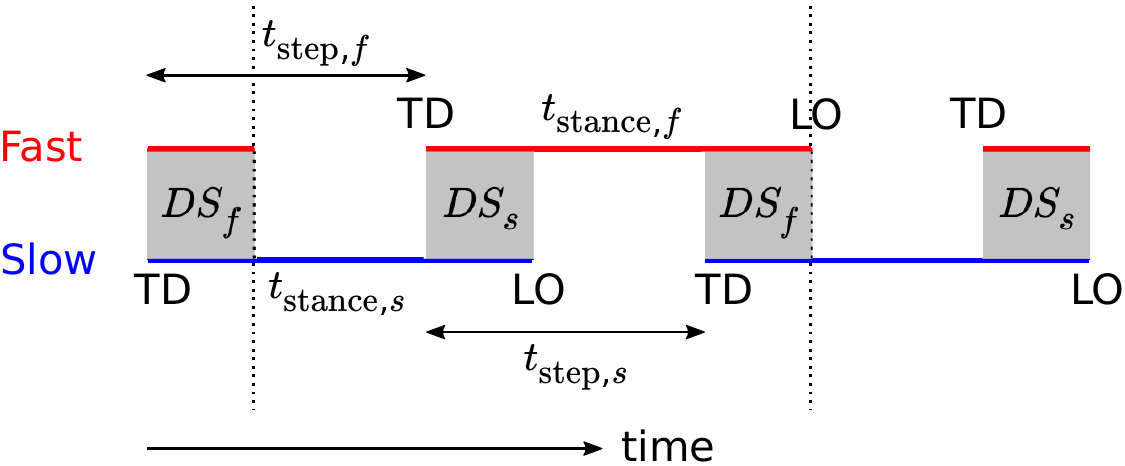}
    \caption{Illustration showing the temporal gait parameters, step time, and stance time. The horizontal direction indicates time, increasing from left to right. A stride, consisting of a step and stance period of each leg, is illustrated by the region between the two dotted lines. The red and blue horizontal lines correspond to periods of contact between the belts and the fast and slow limb, respectively, and the empty areas between the lines of same color correspond to the swing time periods. LO: time of lift-off, TD: touchdown, DS: double support.}
    \label{fig:gait_temporal_parameters}
\end{figure}

For the analysis, the parameters related to the gait are illustrated in Figure \ref{fig:gait_spatial_parameters} and \ref{fig:gait_temporal_parameters} for the spatial and temporal parameters, respectively. These are parameters that can be measured in every stride. We define a stride as one full locomotion cycle, consisting of a step on each limb with double-support periods in between. For the analysis of the gait, the joint angles, toe positions, and ground contact sensor data of the feet is logged. All of the gait parameters are then derived from this data, using the mathematical definitions found in Appendix \ref{app:gait_parameters}.

\subsection{Control architecture} 
\begin{figure}[tb]
\centering
\includegraphics[width=0.9\linewidth]{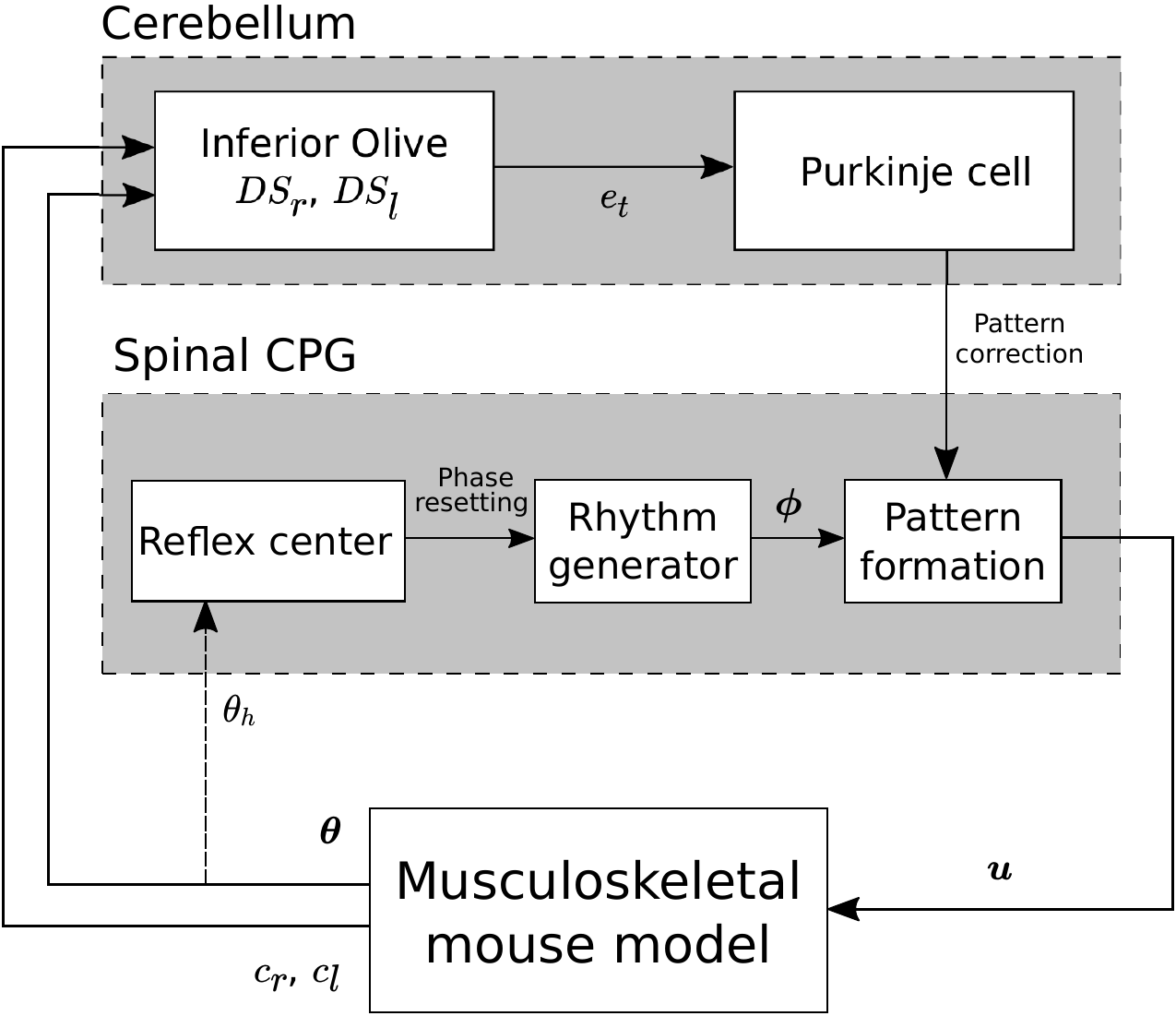}
\caption{The bio-inspired control architecture used for adaptive locomotion of the mouse hindlimbs. The spinal CPG provides timing and patterns of muscle activation $u$ while the cerebellar model provides feed-forward corrections to the step time for symmetric double support times. $\bm\theta$ is the vector of all hindlimb joint angles, $\theta_h$ are the hip angles, and $c_r$ and $c_l$ are the contact sensor values of the right and left hindlimb, respectively.}
\label{fig:control_architecture}
\end{figure}

The control architecture consists of a spinal \gls{cpg} and a cerebellar-like module for adaptation (see Fig. \ref{fig:control_architecture}). The \gls{cpg} is the same as described by Fujiki et al. \cite{Fujiki2018}, but adapted and extended for this mouse model. It is comprised of the rhythm generation and pattern formation layers, with the latter generating the motor neuron activation patterns made of a series of rectangular pulses. In this work, the periodic activation pattern was extended to contain a fourth activation pulse. The addition of a fourth activation pulse was inspired by kinematic data from real mice \cite{Leblond2003}. Here the four gait stages are enumerated as \textit{flexion} (F), \textit{contact} (E1), \textit{extension} (E2), and \textit{push off} (E3). Furthermore, the parameters of the CPG were adjusted to this mouse model.

In the \gls{cpg}, the rhythm is generated by two oscillators, one for each hindlimb, giving a phase $\phi\in[0,2\pi]$ for each. For the rhythm generator parameters, the frequency was chosen to $\omega = \SI{19}{rad/s}$ and the oscillator coupling strength to $K_\phi$ = 7.5. A phase-resetting mechanism keeps the gait in sync with the belt. The phase of each oscillator is reset when the corresponding hip angle reaches a threshold. 

The motor neuron command is formed in the pattern formation layer. It consists of four activation pulses, corresponding to each gait stage. From the phase signal, pulses of unit magnitude are generated by the following equation:
\begin{equation}\label{eq:cpg_pulse}
    u_{j}^{\textnormal{CPG}}(\phi) = \begin{cases} 
      1, & \phi_{j}^\textnormal{start} \leq \phi \leq \phi_{j}^\textnormal{start} + \Delta \phi_j\\
      0, & \textnormal{otherwise}
   \end{cases}, j = {1 \dots 4},
\end{equation}
where $\phi_{j}^\textnormal{start}$ is the pulse phase onset, $\Delta \phi_j$ is the duration, and $j$ denotes the activation pulse, E1, E2, E3, or F.

The motor neuron command $u_m$ is then given as a weighted sum of the pulses, with weighing coefficients $w_{m,j}$:
\begin{equation} \label{eq:cpg_activation}
    u_{m} = \sum_{j=1}^{4} w_{m,j} \cdot u_{j}^{\textnormal{CPG}}(\phi), \; m=1 \dots 8,
\end{equation}
where $m$ denotes each of the muscles.
For more details on the \gls{cpg} model, see \cite{Fujiki2018}.

In the cerebellar microcircuit, the inferior olive (IO) gives the teaching or error signal responsible for driving the adaptation. In the proposed cerebellar-like module, the IO estimates the double support asymmetry from the contact sensor, to give the temporal error signal $e_t$. The asymmetry is given by the difference in the estimated double support times:
\begin{equation}
e_t = DS_s - DS_f,
\end{equation}
where $DS_s$ and $DS_f$ are the slow and fast double support time, respectively. Fast double support time corresponds to the duration of the double support period which ends with liftoff of the limb on the fast belt, and similarly for slow double support time (see Fig. \ref{fig:gait_temporal_parameters}). The double support times are estimated from the contact sensor data using a state machine with timers.

In the cerebellar microcircuit, the  Purkinje cell (PC) is the central learning component. In the proposed cerebellar-like model, the output from the PC is a correction to the duration of the flexion activation pulse in the pattern formation structure of the \gls{cpg}:
\begin{equation}
    \Delta\phi_F = \Delta\phi_F^0 + y_\textnormal{cerebellum},
\end{equation}
where $y_\textnormal{cerebellum}$ is the correction and $\Delta\phi_F^0$ is the baseline value of the flexion pulse duration. This adjusts the duration of the flexion pulse to adjust the step time, and was chosen to allow correction of the double support time in accordance with Figure \ref{fig:gait_temporal_parameters}. The pattern formation part of the CPG was selected as target for the correction because it contains parameters related to the separate stages of the gait, thus the step time is more easily influenced. As the error signal is only updated after each step, a correction to the rhythm generation part would be constant in one locomotion cycle, and thus affect all stages of the gait equally.

To ensure that the flexion (F) pulse and the early extension (E1) pulse have the same phase spacing, the cerebellum also corrects the onset phase of the E1 pulse:
\begin{equation}
    \phi_{E1}^{Start} = \phi_{E1}^{Start,0} + y_\textnormal{cerebellum},
\end{equation}
where $\phi_{E1}^{Start,0}$ is the baseline value for the onset phase of the E1 pulse.

The cerebellar learning rule is a gradient-descent type with learning rate $\alpha^*$ which is applied at the end of every period of double support. The ($i+1$)'th output values for each limb is given by
\begin{equation}
    y_\textnormal{cerebellum, s}^{(i+1)} = y_\textnormal{cerebellum}^{(i)} - \alpha^* e_t,
\end{equation}
and
\begin{equation}
    y_\textnormal{cerebellum, f}^{(i+1)} = y_\textnormal{cerebellum}^{(i)} + \alpha^* e_t,
\end{equation}
for the slow and fast limb, respectively. The derivation of the signs for the update rules can be found in Appendix \ref{app:learning_rule}.

\section{Results}\label{sec:results}
\begin{figure*}[htb]
    \centering
    \centering
    \includegraphics[width=0.2\linewidth]{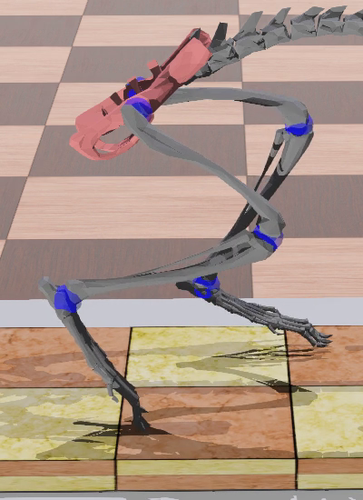}%
    \includegraphics[width=0.2\linewidth]{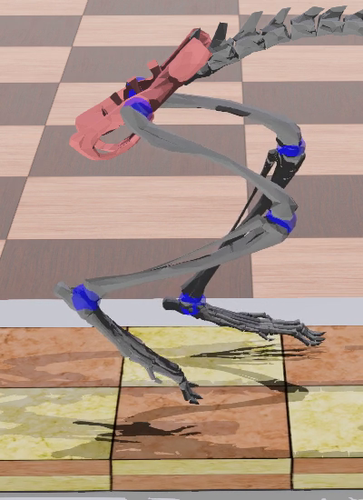}%
    \includegraphics[width=0.2\linewidth]{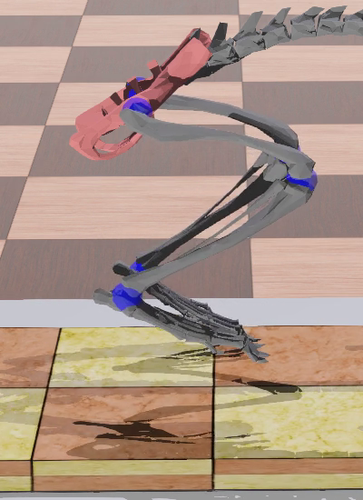}%
    \includegraphics[width=0.2\linewidth]{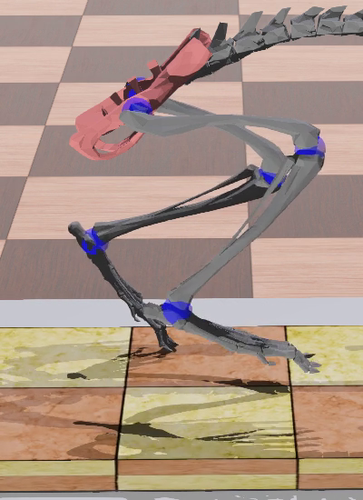}%
    \includegraphics[width=0.2\linewidth]{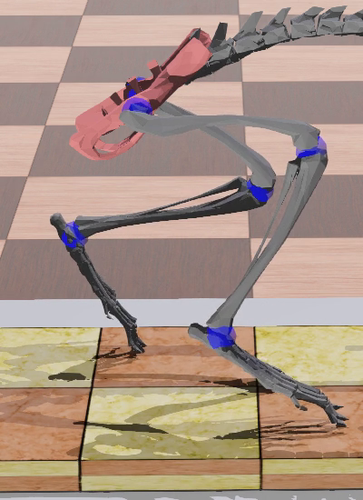}%
    
    \includegraphics[width=0.2\linewidth]{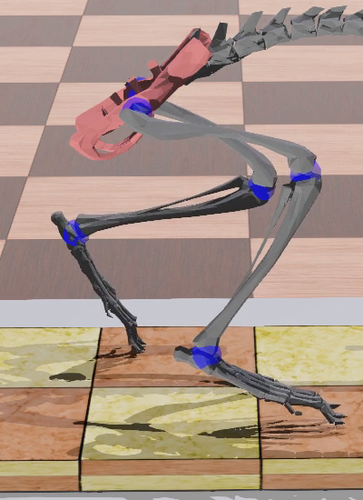}%
    \includegraphics[width=0.2\linewidth]{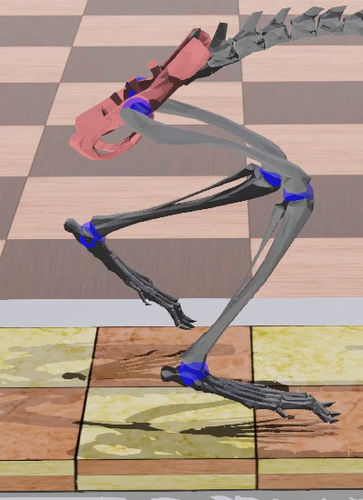}%
    \includegraphics[width=0.2\linewidth]{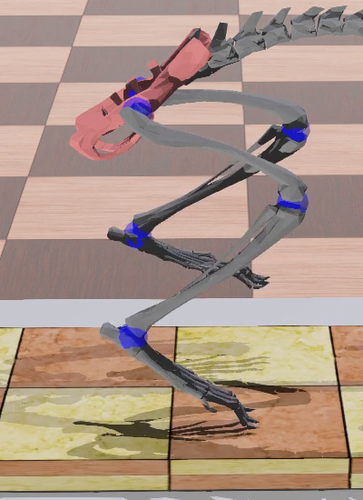}%
    \includegraphics[width=0.2\linewidth]{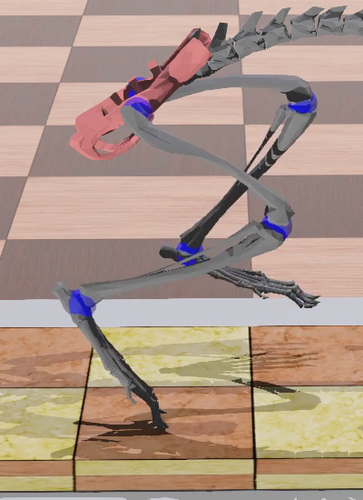}%
    \includegraphics[width=0.2\linewidth]{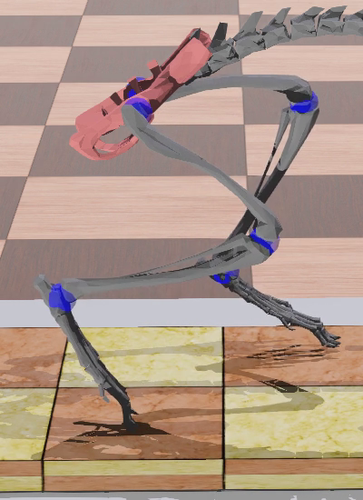}%
    \caption{Series of images showing the base gait on a single belt. The locomotion direction is to the right, thus the belt moves to the left.}\label{fig:gait_video_pictures}
\end{figure*}

The gait on a single treadmill is shown in Figure \ref{fig:gait_video_pictures} where the control architecture has been applied without the cerebellar model.

\begin{figure}[bh]
    \centering
    \begin{subfigure}[b]{.49\linewidth}
        \centering
        \includegraphics[width=\linewidth]{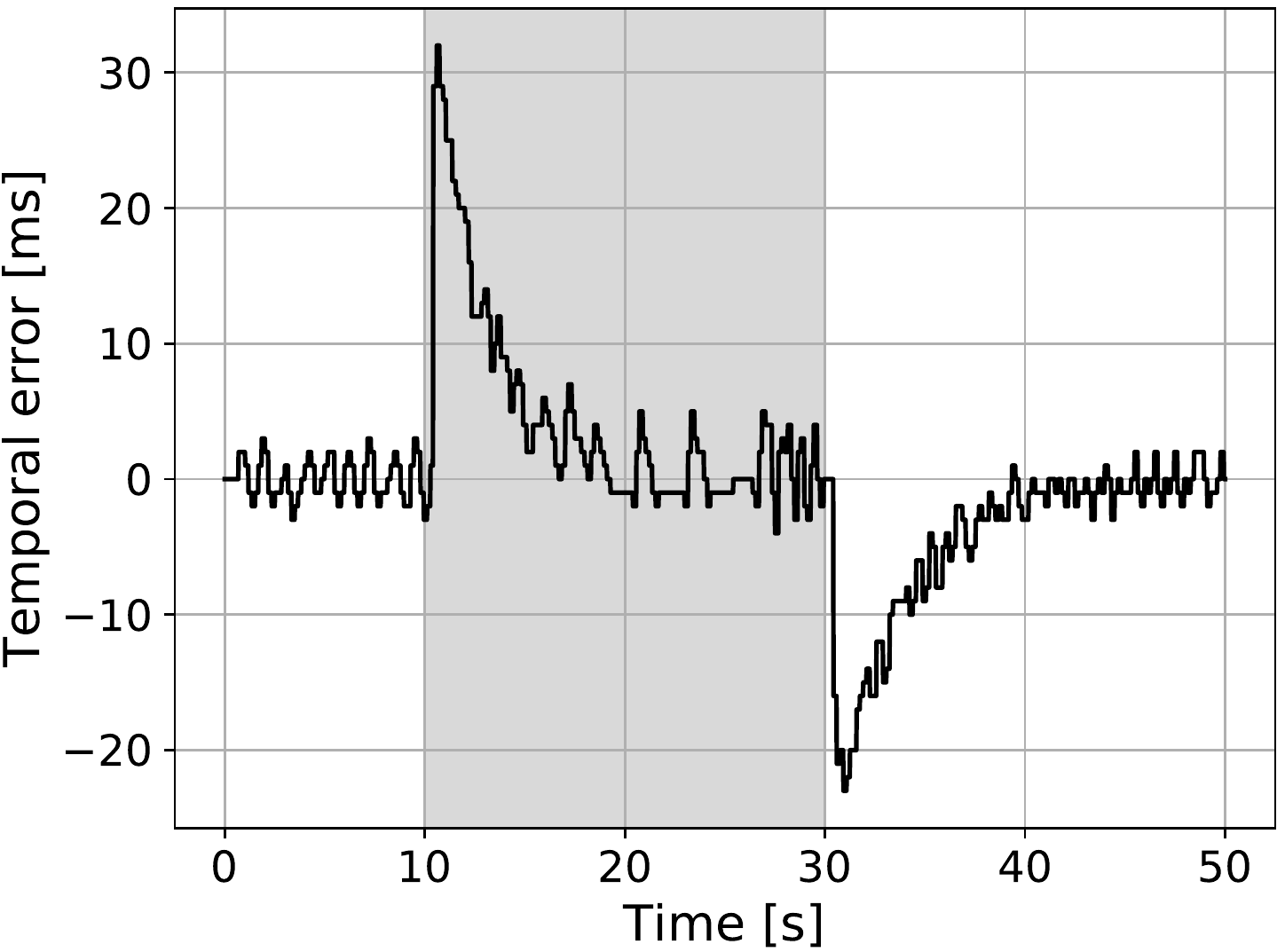}
        \caption{Error.}
        \label{fig:online_temporal_error}
    \end{subfigure}
    \hfill
    \begin{subfigure}[b]{.49\linewidth}
        \centering
        \includegraphics[width=\linewidth]{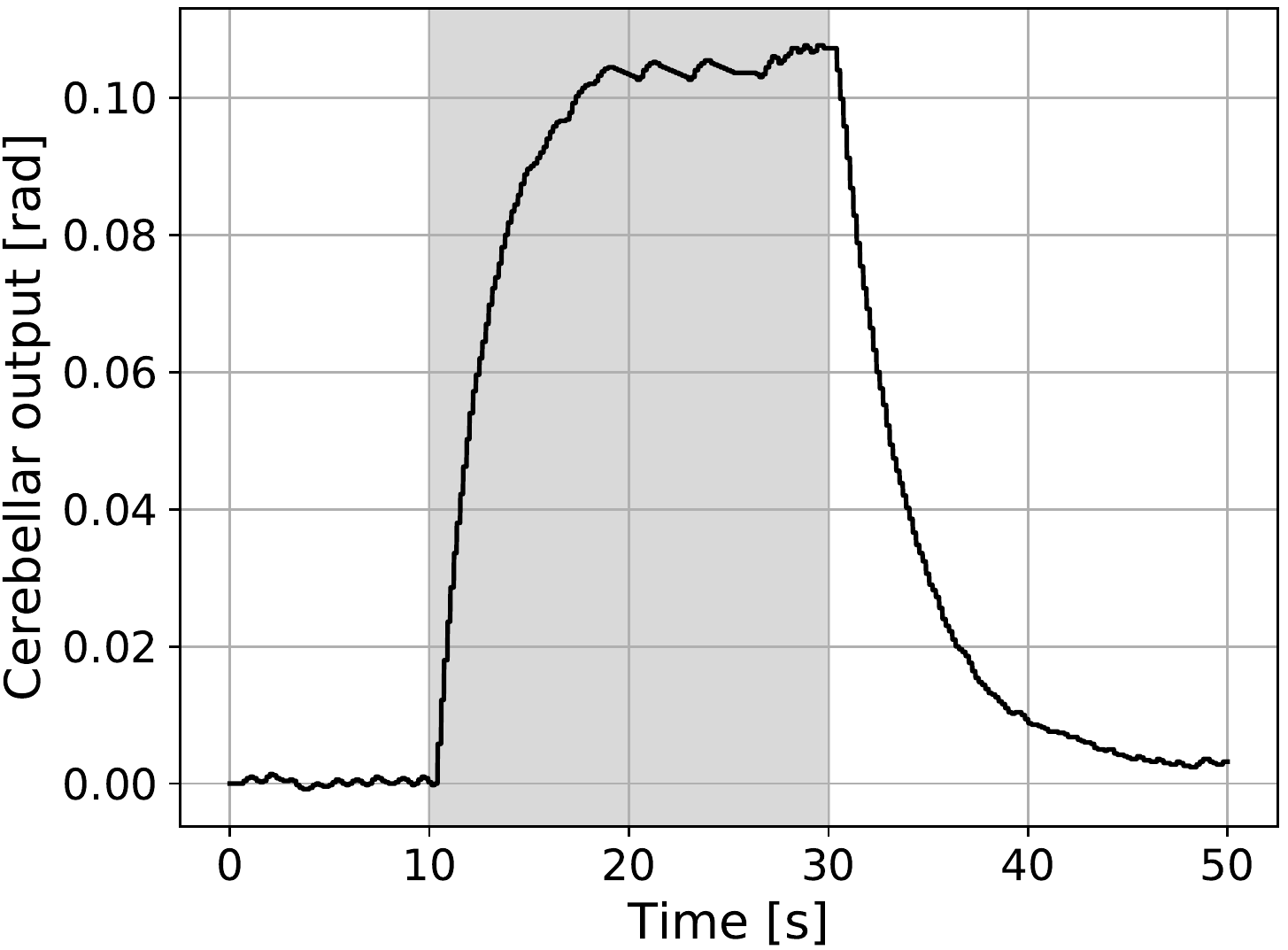}
        \caption{Output.}
        \label{fig:online_cerebellum_output}
    \end{subfigure}
    \caption{The plots show the $e_t$ online error estimate and the $y_{\textnormal{cerebellum,f}}$ output of the cerebellar-like model for the fast limb in the split-belt protocol with a belt velocity ratio of 1.5. The belt speed is perturbed at $t=10 s$ as shown by the grayed out region.}
    \label{fig:cerebellar_learning}
\end{figure}

We now apply the full control architecture in the split-belt trial. Figure \ref{fig:cerebellar_learning} shows that the cerebellar-like adaptation is able to reduce the double support asymmetry in the split-belt protocol by making corrections to the pattern formation part of the \gls{cpg}. As the output signals are symmetric around zero, only the output for the fast limb, $y_{\textnormal{cerebellum,f}}$, is shown.

\begin{figure}[tb]
    \centering
    \begin{subfigure}[b]{.49\linewidth}
        \centering
        \includegraphics[width=\linewidth]{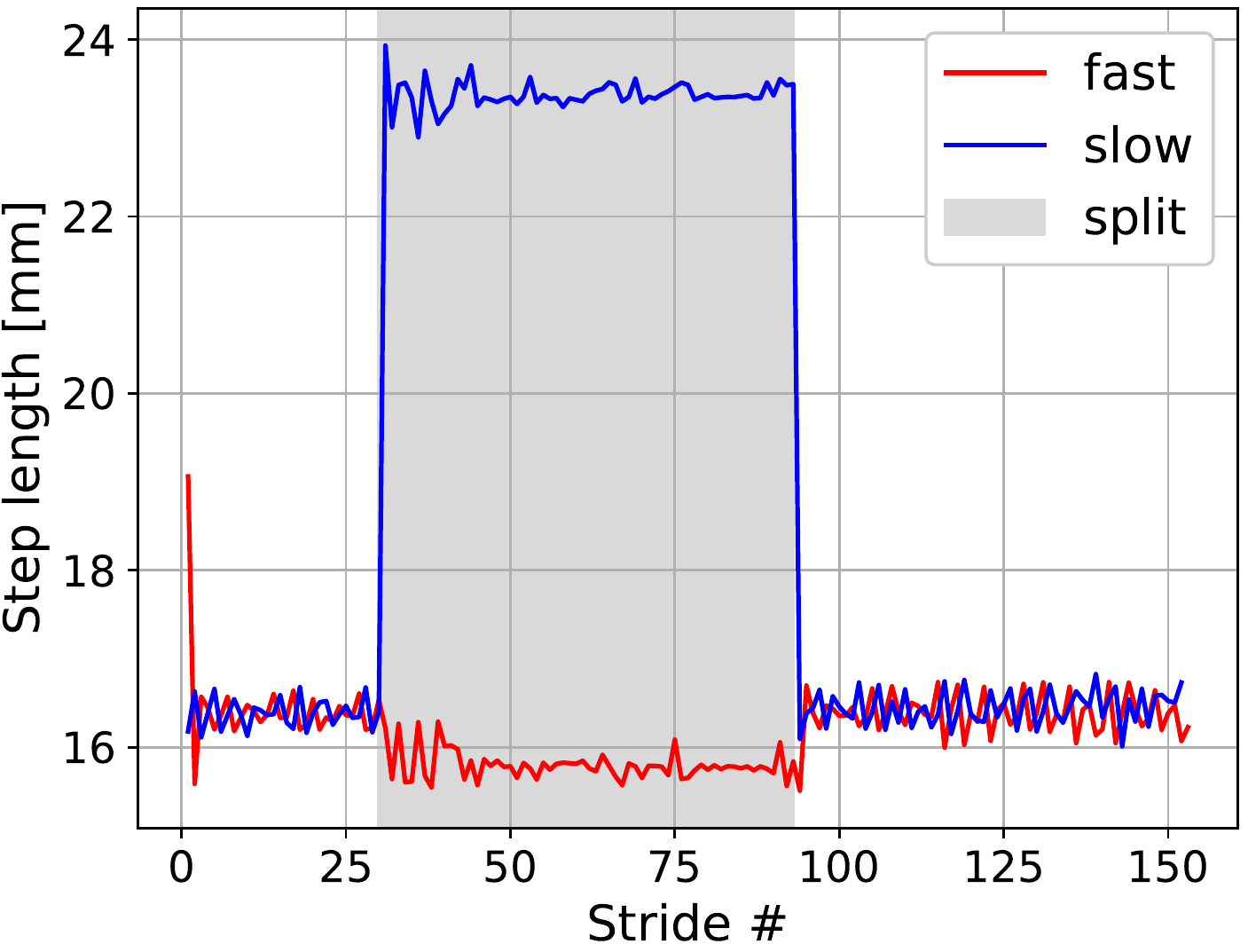}
        \caption{Without cerebellum.}
        \label{fig:step_length_baseline}
    \end{subfigure}
    \hfill
    \begin{subfigure}[b]{.49\linewidth}
        \centering
        \includegraphics[width=\linewidth]{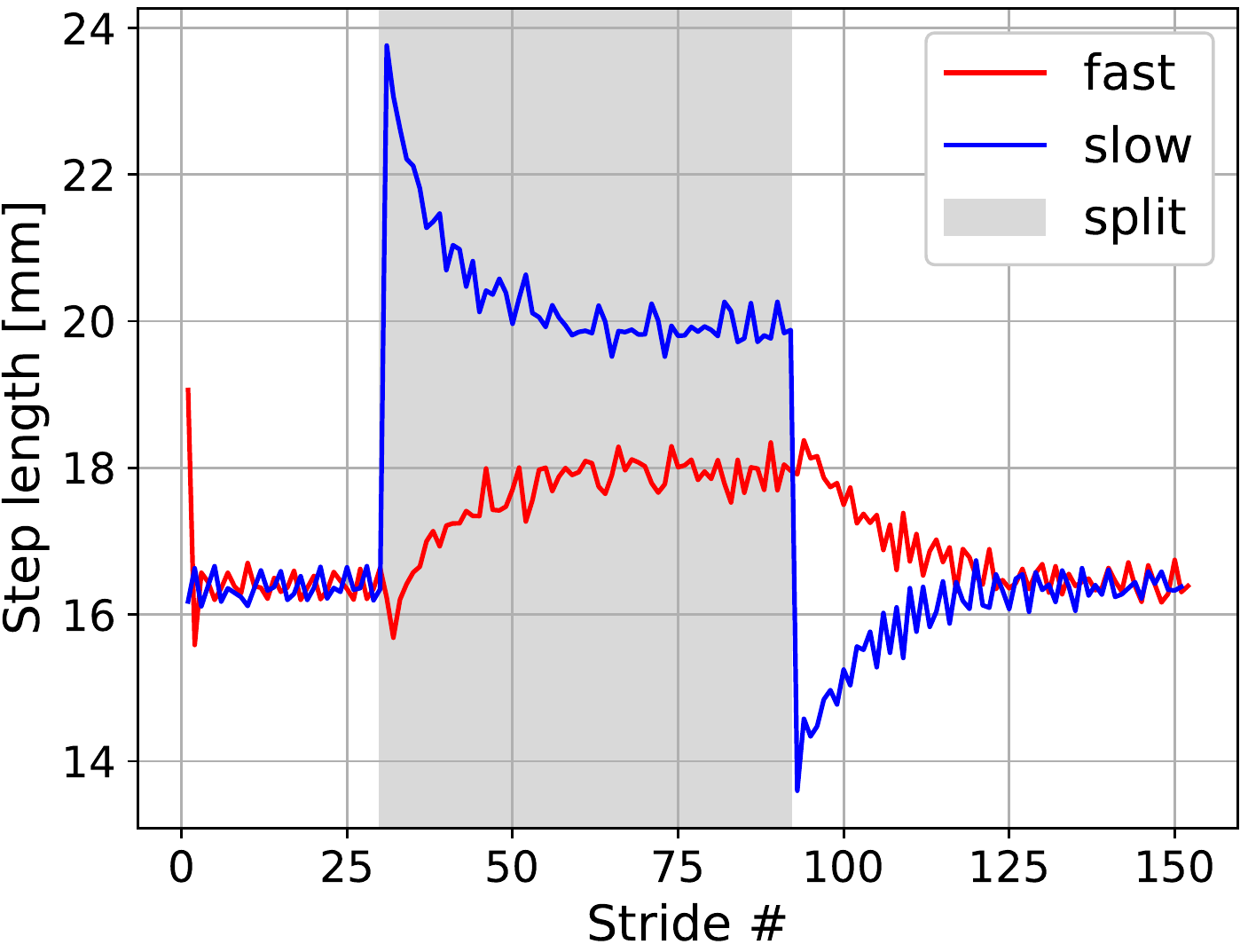}
        \caption{With cerebellum.}
        \label{fig:step_length_learning}
    \end{subfigure}
    \caption{Step length in the split-belt trial with a 1.5x speed ratio.}
    \label{fig:step_lengths}
\end{figure}

\begin{figure}[tb]
    \centering
    \begin{subfigure}[b]{.49\linewidth}
        \centering
        \includegraphics[width=\linewidth]{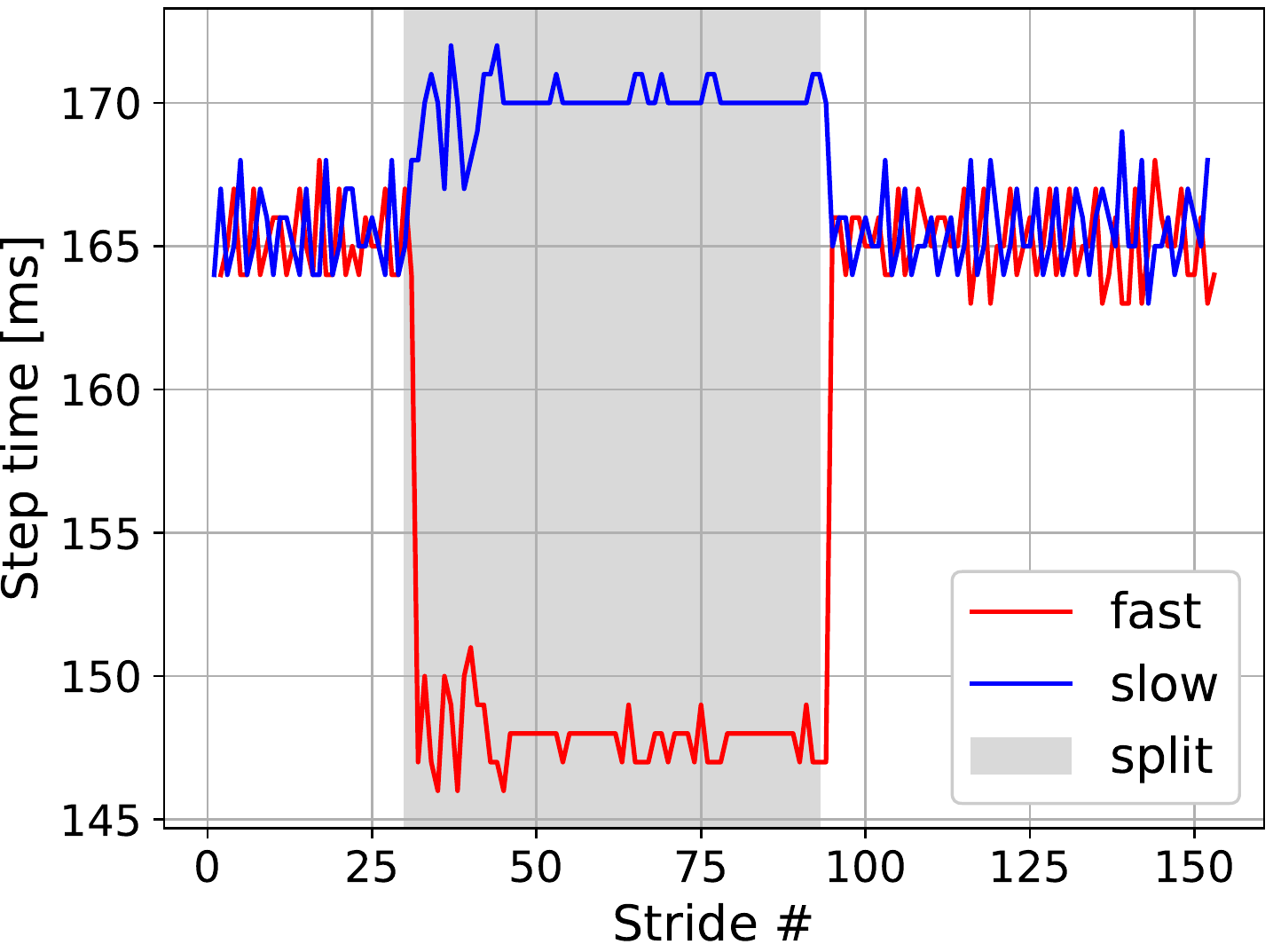}
        \caption{Without cerebellum.}
        \label{fig:step_time_baseline}
    \end{subfigure}
    \hfill
    \begin{subfigure}[b]{.49\linewidth}
        \centering
        \includegraphics[width=\linewidth]{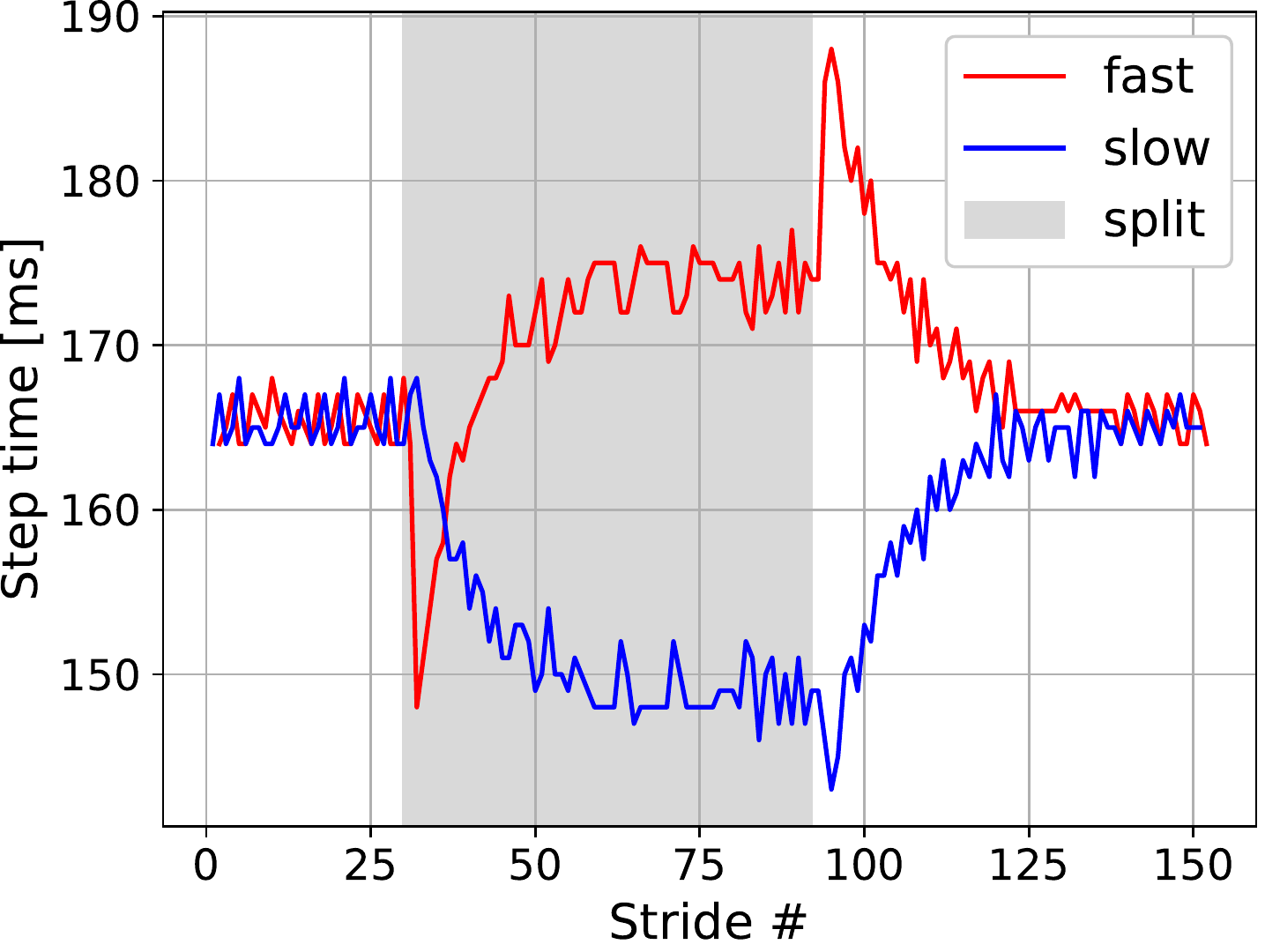}
        \caption{With cerebellum.}
        \label{fig:step_time_learning}
    \end{subfigure}
    \caption{Step time in the split-belt trial with a 1.5x speed ratio.}
    \label{fig:step_times}
\end{figure}

\begin{figure}[tb]
    \centering
    \begin{subfigure}[b]{.49\linewidth}
        \centering
        \includegraphics[width=\linewidth]{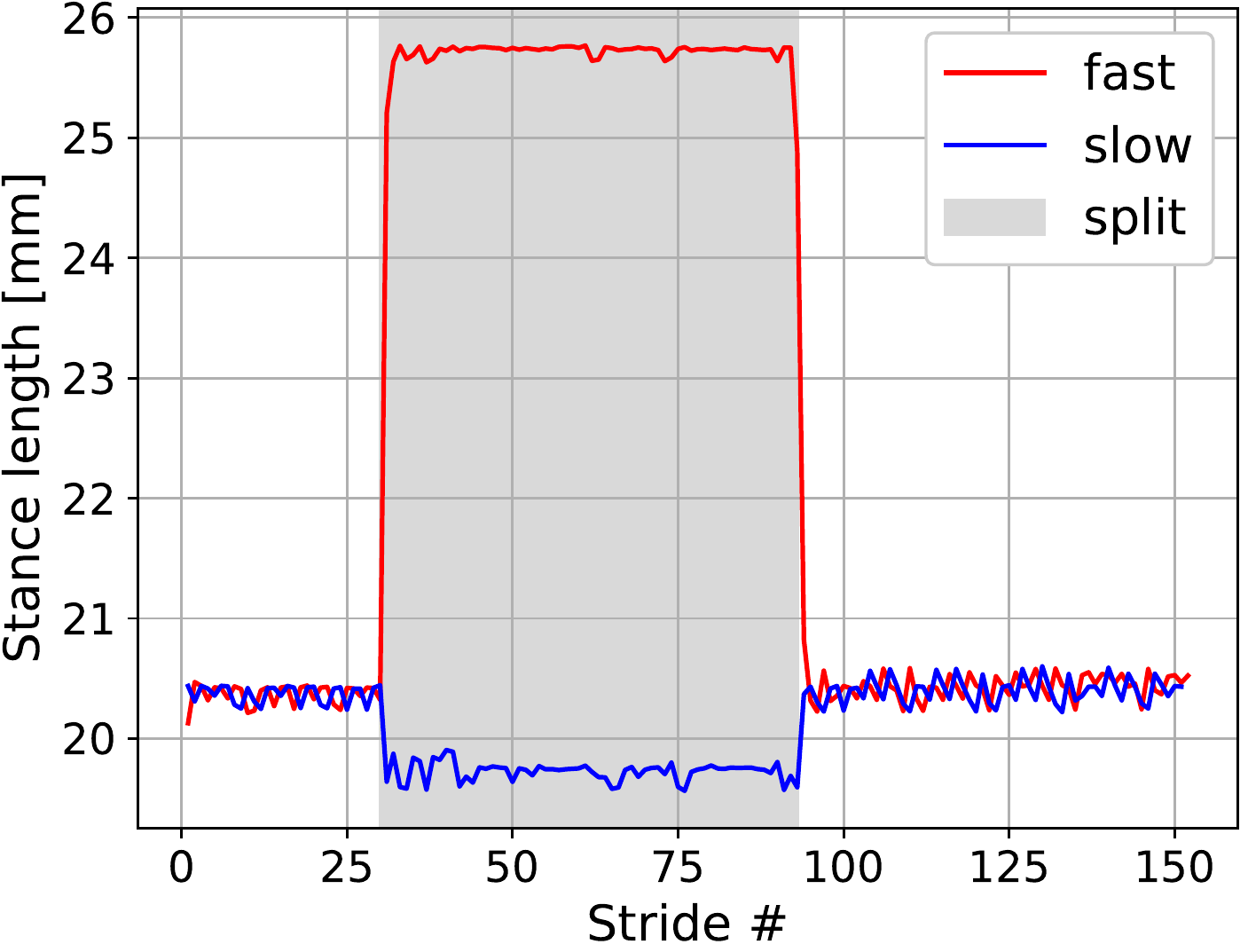}
        \caption{Without cerebellum.}
        \label{fig:stance_length_baseline}
    \end{subfigure}
    \hfill
    \begin{subfigure}[b]{.49\linewidth}
        \centering
        \includegraphics[width=\linewidth]{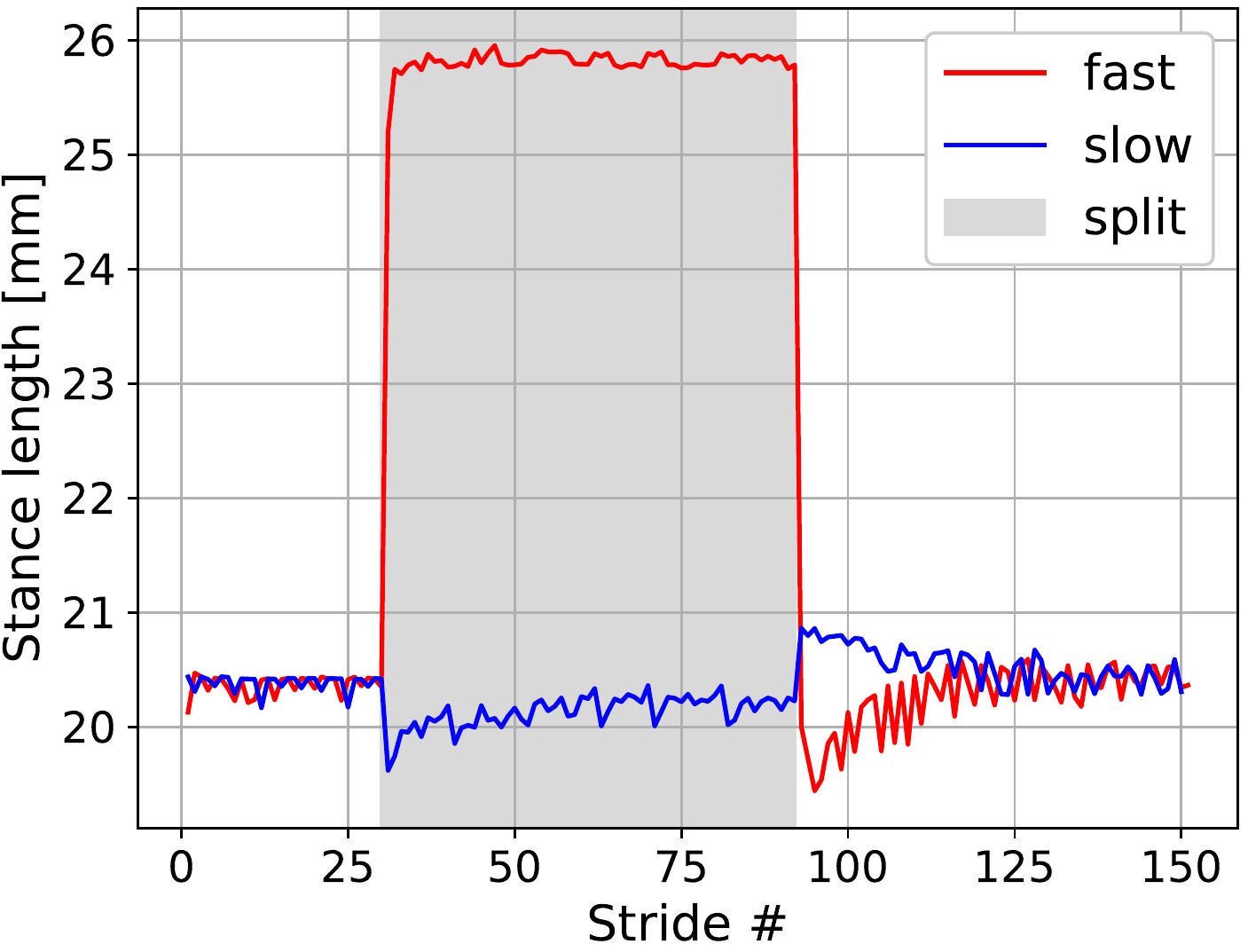}
        \caption{With cerebellum.}
        \label{fig:stance_length_learning}
    \end{subfigure}
    \caption{Stance length in the split-belt trial with a 1.5x speed ratio.}
    \label{fig:stance_lengths}
\end{figure}

\begin{figure}[tb]
    \centering
    \begin{subfigure}[b]{.49\linewidth}
        \centering
        \includegraphics[width=\linewidth]{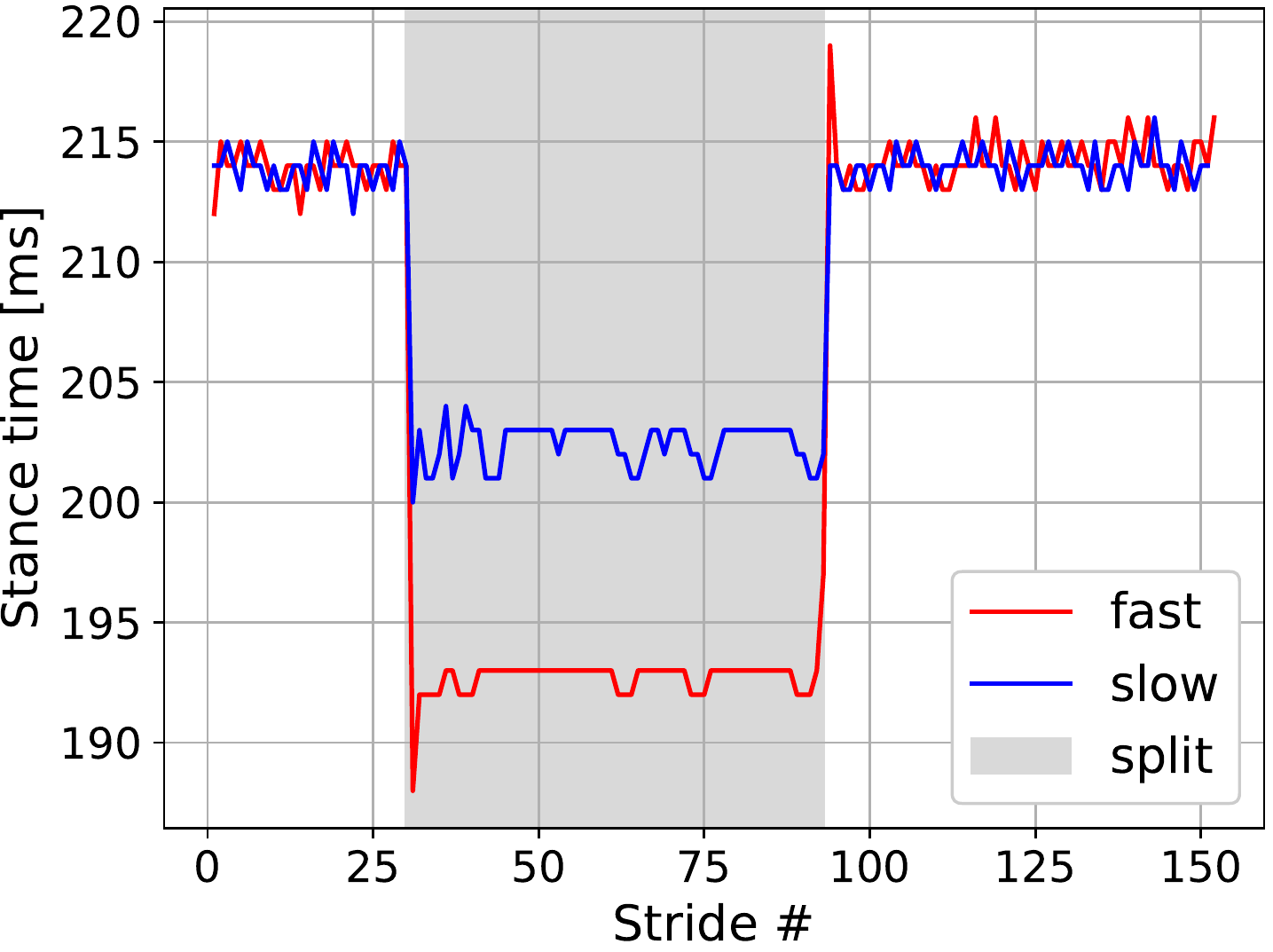}
        \caption{Without cerebellum.}
        \label{fig:stance_time_baseline}
    \end{subfigure}
    \hfill
    \begin{subfigure}[b]{.49\linewidth}
        \centering
        \includegraphics[width=\linewidth]{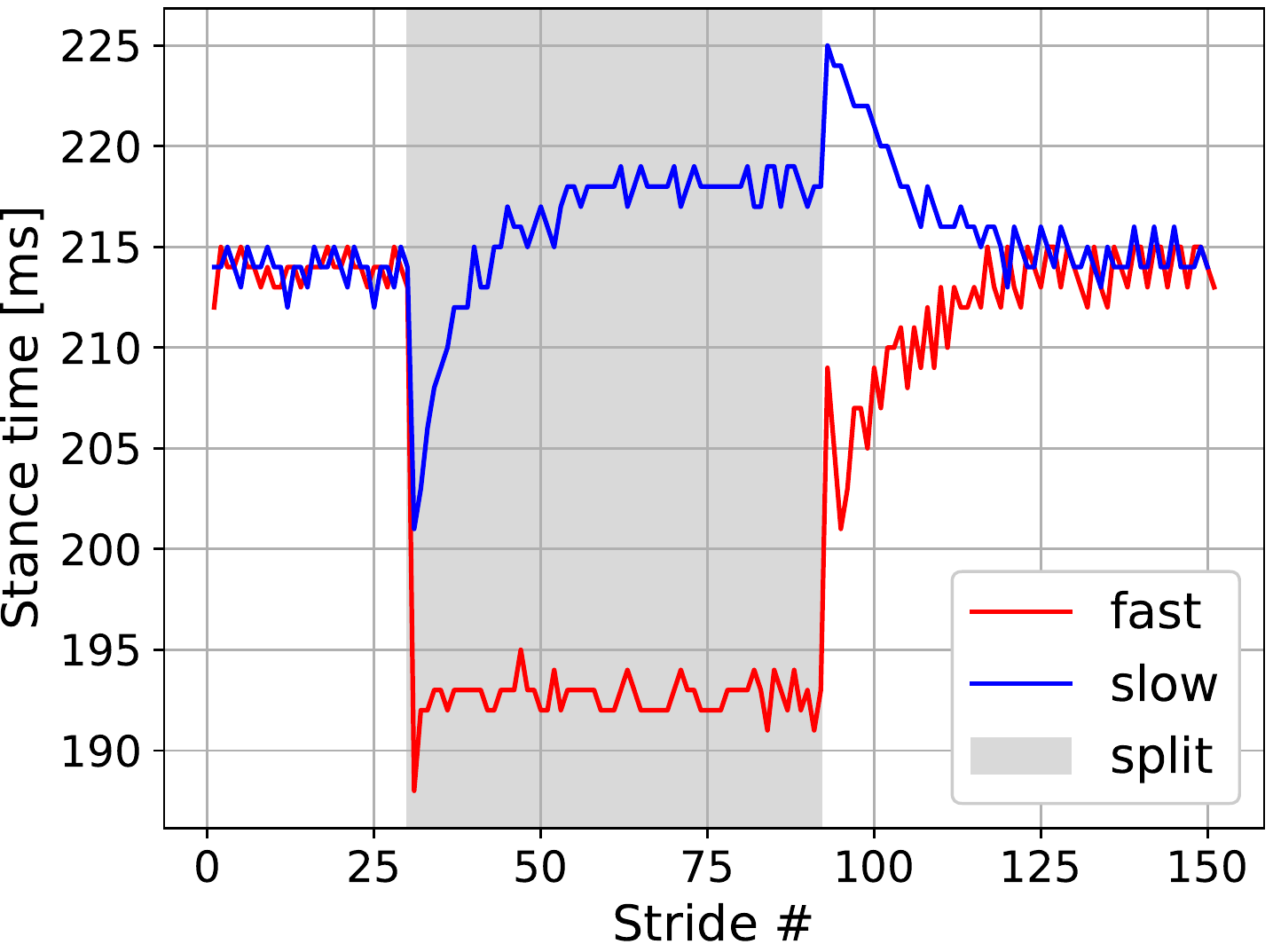}
        \caption{With cerebellum.}
        \label{fig:stance_time_learning}
    \end{subfigure}
    \caption{Stance time in the split-belt trial with a 1.5x speed ratio.}
    \label{fig:stance_times}
\end{figure}

\begin{figure}[tb]
    \centering
    \begin{subfigure}[b]{.49\linewidth}
        \centering
        \includegraphics[width=\linewidth]{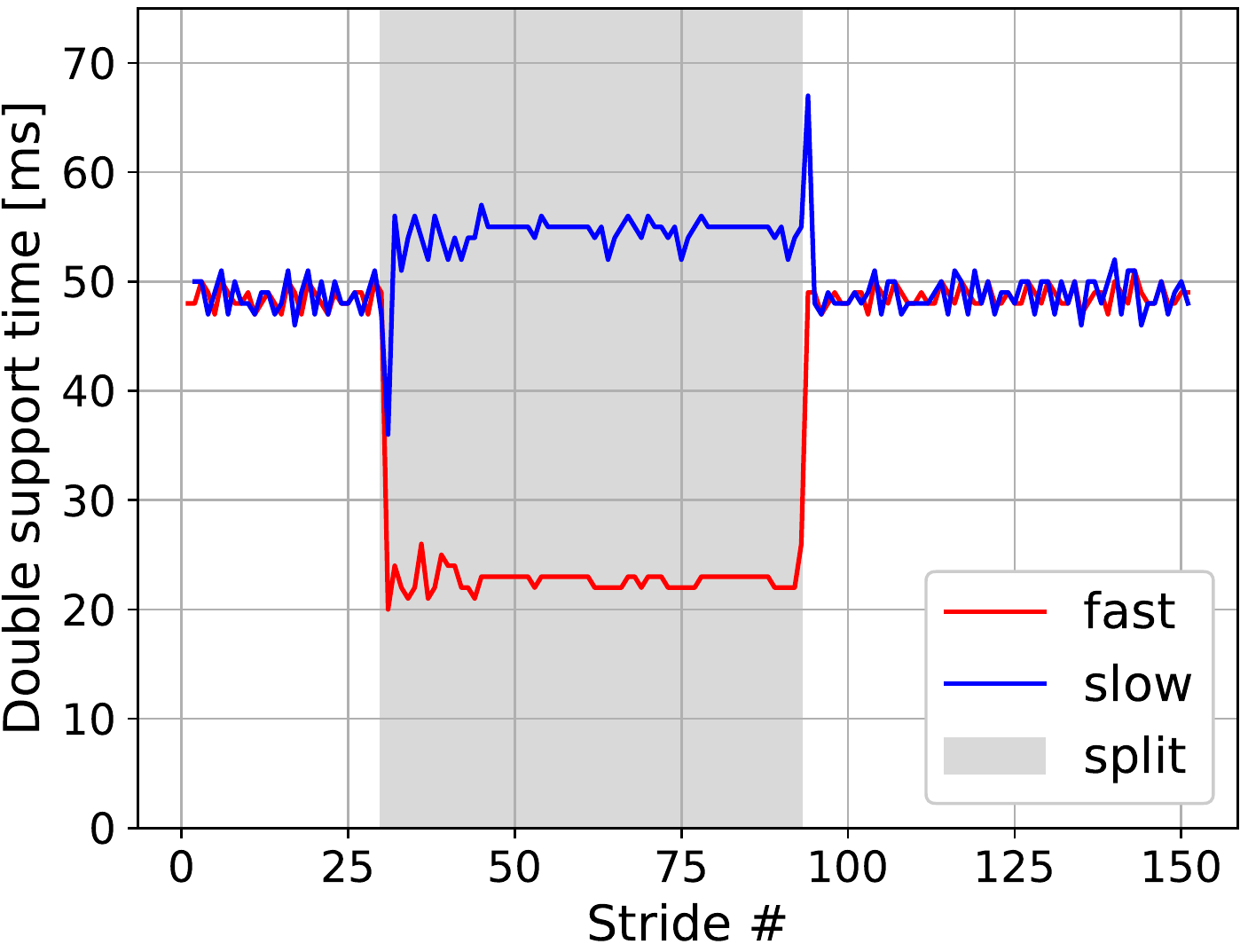}
        \caption{Without cerebellum.}
        \label{fig:double_support_time_baseline}
    \end{subfigure}
    \hfill
    \begin{subfigure}[b]{.49\linewidth}
        \centering
        \includegraphics[width=\linewidth]{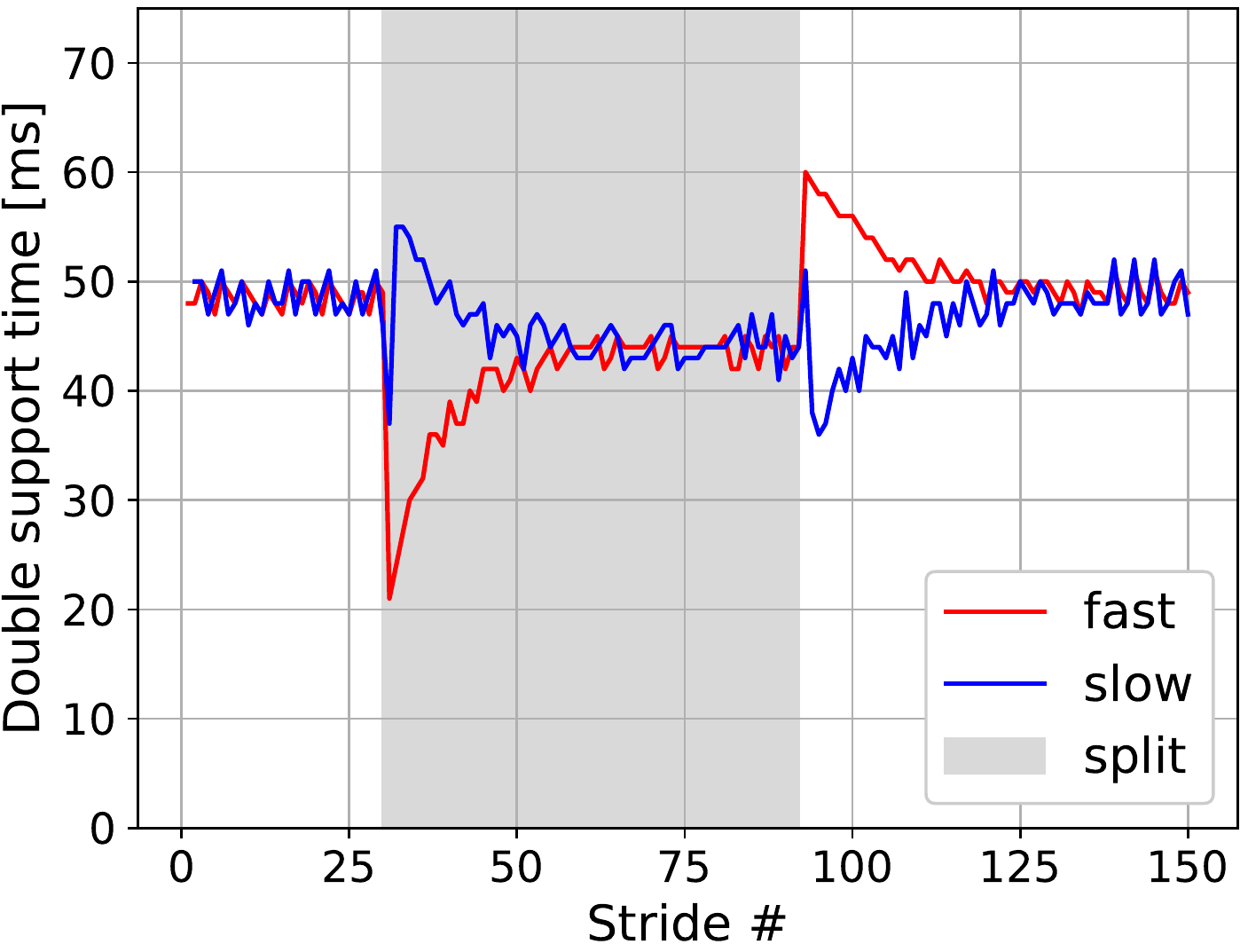}
        \caption{With cerebellum.}
        \label{fig:double_support_time_learning}
    \end{subfigure}
    \caption{Double support time in the split-belt trial with a 1.5x speed ratio.}
    \label{fig:double_support_times}
\end{figure}

The experimental protocol was carried out with and without cerebellar learning. For the experiment with a velocity ratio of 1.5, the time-evolution of the temporal and spatial parameters presented in Section II, throughout the protocol, are compared for the mouse with and without cerebellar learning in Figures \ref{fig:step_lengths} to \ref{fig:double_support_times}. %

It is seen in Figure \ref{fig:step_time_learning} that the fast step time is increased as the cerebellar output for that limb is increased (which is not surprising given the derivation in Appendix  \ref{app:learning_rule}). However, the remainder of the gait parameters also changes, as a result thereof. When the belt speed is perturbed, all of the gait parameters change immediately in the early adaptation period. Without cerebellar learning, the gait parameters still change due to the hip reflex, which attempts to keep up to the faster belt speed, but they are not adapted during the early- and late adaptation periods. This reflex effect happens already at the first stride after the perturbation and does not change after that. Without learning, there is a constant double support asymmetry (Fig. \ref{fig:double_support_time_baseline}) during the perturbation, which returns to zero immediately when the perturbation is removed in post-adaptation.

The cerebellar adaptation does not only affect the step time, but also the remaining gait parameters to a varying degree. While no effect is seen on the stance length (Fig. \ref{fig:stance_lengths}) and stance time (Fig. \ref{fig:stance_times}) of the fast leg at all, the stance length and time for the slow leg increases during the adaptation. When the step time of the slow limb is decreased, the stance time automatically increases because the total stride time is approximately constant. The total stride time increases marginally when the effective oscillator frequency increases slightly, due to the hip reflex. The reason for no increase in the fast stance length could be that the hip reflex is activating, initiating the lift-off of the limb.

The slow stance time is increased by a considerable amount (Fig. \ref{fig:stance_time_learning}), which is working against the adaptation. To decrease the double support asymmetry, the cerebellum is shortening the slow double support time and lengthening the fast double support time. To shorten the slow double support time, the fast step time is increased (see Fig. \ref{fig:gait_temporal_parameters}). But Figure \ref{fig:stance_time_learning} shows that the slow stance time is increasing, which is working against decreasing the slow double support time.

The step length (Fig. \ref{fig:step_length_learning}) seems directly correlated to step time, as one would expect; when the step time is increased due to the adaptation, the step length increases as well.

With cerebellar learning, the double support asymmetry is reduced from the early to late adaptation period, but when the perturbation is removed, a double support asymmetry with the opposite sign arises in post-adaptation (Fig. \ref{fig:online_temporal_error}). In late post-adaptation, this error has been reduced by the learning rule. This after-effect happens because the cerebellar output has to adapt back to the baseline value, as seen in Figure \ref{fig:online_cerebellum_output}.


\begin{figure}[tb]
    \centering
    \begin{subfigure}[b]{.49\linewidth}
        \centering
        \includegraphics[width=\linewidth]{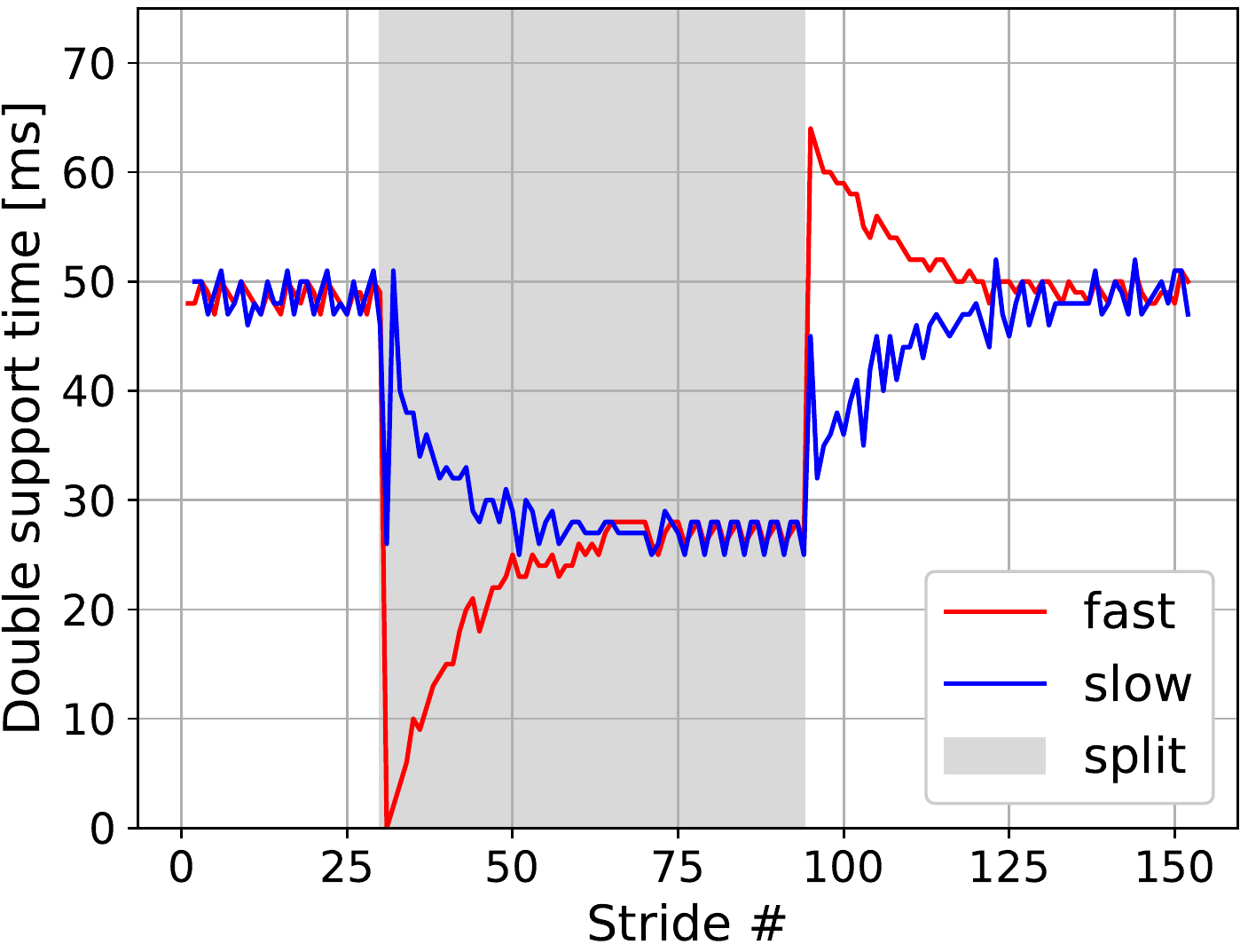}
        \caption{1.7x}
        \label{fig:double_support_time_1.7}
    \end{subfigure}
    \hfill
    \begin{subfigure}[b]{.49\linewidth}
        \centering
        \includegraphics[width=\linewidth]{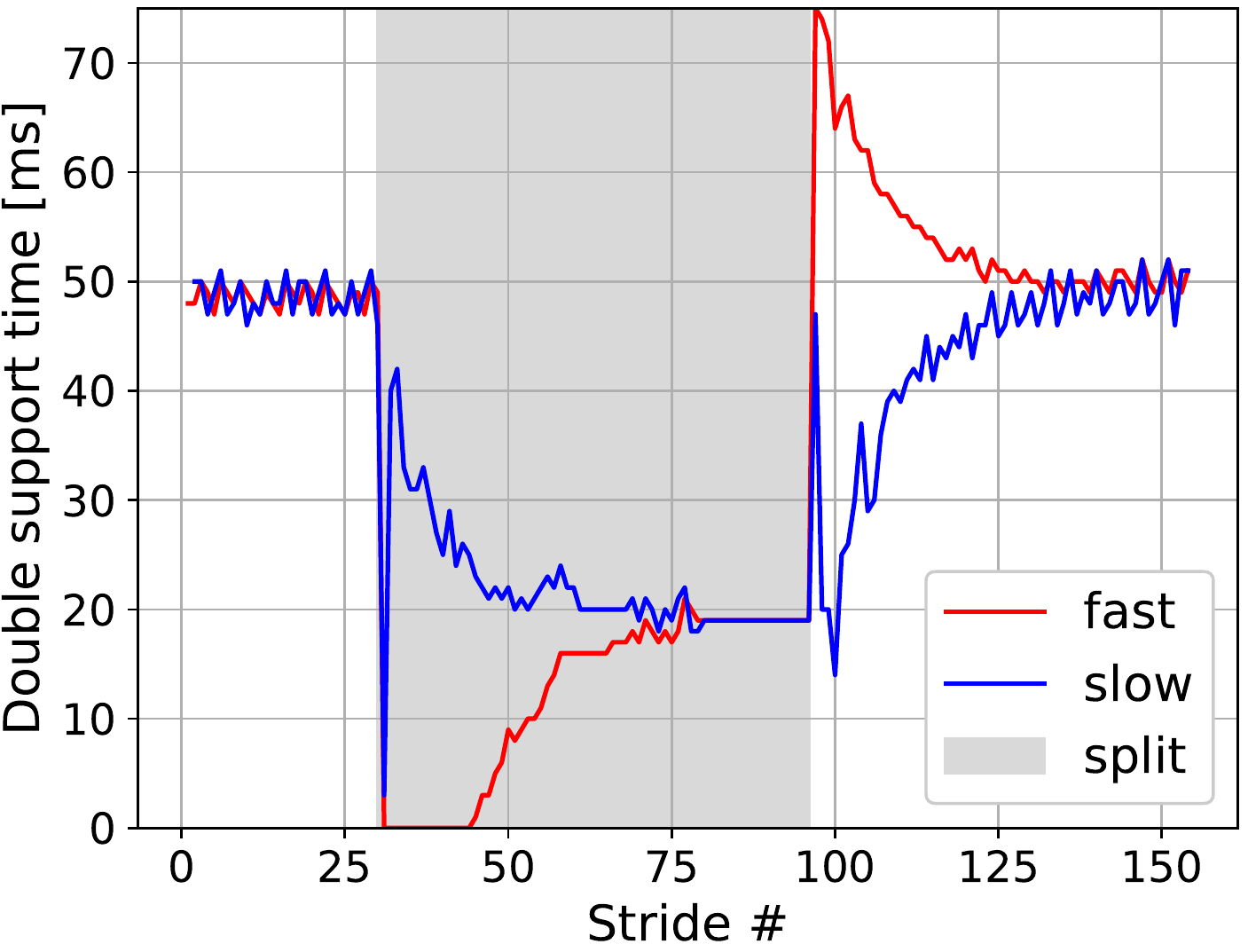}
        \caption{2.0x}
        \label{fig:double_support_time_2.0}
    \end{subfigure}
    \caption{Double support time in the split-belt trial with two higher belt velocity ratios.}
    \label{fig:double_support_time_factors}
\end{figure}

The double support adaptation is shown in Figure \ref{fig:double_support_time_learning} and \ref{fig:double_support_time_factors} for each of the relative belt velocities (1.5x, 1.7x, and 2.0x). It can be seen that the initial asymmetry is larger for higher velocity ratios, and for 2.0x there is initially no period of fast double support. Still, in that case, the cerebellar adaptation is able to reduce the asymmetry. When adapting, both slow and fast double support times are changing, and the final late-adaptation value is lower for higher velocity ratios. When the fast limb has to take longer steps due to the faster moving belt, more time is spent in the swing stage, and there is less time for double support. The adaptation time is longer for higher velocity ratios, having adapted after $\approx$30, $\approx$40, and $\approx$50 strides. 
The magnitude of the after-effect and the time to de-adapt is also larger with a higher velocity ratio. 

\section{Discussion}\label{sec:discussion}
This paper focuses on the input and output of the cerebellum rather than on the evaluation of its learning mechanism. The cerebellar learning rule was derived based on the internal mechanism of the \gls{cpg} model, which enabled the cerebellum to efficiently reduce the double support time asymmetry. This shows that the assumption used to derive the learning rule (see Appendix \ref{app:learning_rule}) is coherent. %
The cerebellar-like module does not capture the complexity of the real cerebellum, however, it is used as a phenomenological model to determine the effect of applying the selected input (double support asymmetry) and output (pattern formation part of the \gls{cpg}) for the cerebellar-like model. We will now compare these results with the ones from the literature, to determine whether the chosen input and output gives rise to similar adaptive behavior.

Fujiki et al. \cite{Fujiki2018} ran simulations of hindlimb locomotion of a rat and compared their results with data from real rats, to determine the importance of phase resetting by a hip reflex in a split-belt trial. The \gls{cpg} model implemented here is a further development of the one from \cite{Fujiki2018}. By comparing our results with theirs, without the cerebellar model, it is clear that the effects of the reflex are qualitatively the same, in that the gait parameters change in same direction due to the split-belt perturbation. This hip reflex, or one with a similar effect, is very important for a stable gait when the belt speed is perturbed. %

The comparison of mice walking on hindlimbs and human bipedal locomotion in split-belt trials gives many interesting insights. After all, the only difference is that mice do not have to keep balance, where humans do (although in most experiments they are assisted by a handrail). The results in \cite{Morton2006} correspond very well with the results we obtained by applying the simplistic control architecture on the musculoskeletal model. For example, we also found that the faster reactive adaptations were provided by the reflexes. The correction that is seen in the stance length (Fig. \ref{fig:stance_lengths}) is also longer on the fast limb, and slightly shorter on the slow limb in their case, however, the slow stance time does not show any adaptation in the measurements made by Morton et al. \cite{Morton2006}. The reactive change in the step length in early adaptation is in the same direction, and the adaptation decreases the asymmetry. For the healthy subjects, the step length asymmetry was almost completely removed through the adaptation. The magnitudes of the gait parameters are not expected to be comparable between mice and humans.
In \cite{Reisman2005}, healthy human subjects are subjected to the split-belt, where the results show that the adaptations take more time, for a larger speed ratio. The same is the case here, as seen in Figure \ref{fig:double_support_time_learning} and \ref{fig:double_support_time_factors}.

In \cite{Darmohray2019} it was shown that cerebellar adaptations in mice can be separated into spatial and temporal contributions, with the temporal contribution adapting on a shorter time scale. Also, it was suggested that the adaptation is primarily learned for the forelimbs. Their results showed that the double support asymmetry is reduced in the split-belt period, albeit only for the front limbs. In this paper, only the temporal component was modeled and the simulated mouse is only walking on the hindlimbs. Thus, the results here cannot be compared directly with the ones from the aforementioned paper, however, the adaptive behavior carries a strong resemblance. When comparing the results presented here for the hindlimbs, with their forelimb results, a lot of similarities can be seen. The reactive change in stance length is in their results almost completely symmetrical around the baseline value, which is not the case here, however, both cases show no adaption of stance length in the split period. At the same time, the step length shows very similar adaptation, but the reactive change is not the same. The differences are not surprising when comparing fore- and hindlimbs. In general, the comparison with results from the literature shows a good agreement in that interlimb parameters (step length and duration) are predominantly affected by the cerebellar adaptation while the intralimb parameters (stance length and duration) are almost unaffected.

From the results, it is clear that applying the double support asymmetry as an error signal for adapting step timing in the pattern formation part of the \gls{cpg} does result in adaptive behavior that is similar to that observed for mice and humans. It is unclear whether the real cerebellum does use double support asymmetry as an error signal, or whether a reduction of the double support asymmetry is only an effect of optimization of a higher objective, for example to keep balance \cite{Houk2001}, or to save energy \cite{Finley2013}.

\section{Conclusion}
In this paper, we presented a bio-inspired adaptive control architecture for hindlimb locomotion in the mouse to test the hypothesis that the cerebellum gives corrections to the pattern formation part of the \gls{cpg}, using the double support asymmetry as a temporal teaching signal. 
In our simulated split-belt experiment the musculoskeletal system of a mouse is able to adapt to perturbations coming from the unequal belt velocities, by adjusting the step timing. The comparison with results from real mice and humans in the literature shows that our control architecture captures many of the same adaptive features. %
This contributes to validate the double support asymmetry as the teaching signal used for gait adaptation in humans and mice. At the same time, it gives confidence to the cerebellar output being applied to different parts of the CPG network, and not only as a correction to the motor neuron output.
However, it is clear that in order to capture all of the dynamics, a model of spatial adaptation is needed. For this reason, we hypothesize that cerebellar internal models of limb dynamics can be used to accurately achieve desired foot positions to reduce a spatial error signal, such as the asymmetry in the center of oscillation.

In a future work, it will be of interest to further develop the cerebellar-like model for adaptive quadrupedal locomotion in the mouse, and to investigate if both spatial and temporal errors will allow the musculoskeletal system to keep the balance.

\bibliographystyle{ACM-Reference-Format}
\bibliography{references}

\appendix
\section{Gait parameters}\label{app:gait_parameters}
All gait parameters are defined in terms of the slow or fast limb, where the slow limb refers to the limb on the slow belt, and vice versa for the fast limb.

We define by $x_s(t)$ and $x_f(t)$ the horizontal toe position of the slow and fast limb with respect to time. The stance length is the intralimb distance between the toe positions of touchdown and liftoff. Given as, for the fast limb:
\begin{equation} 
s_{\textnormal{stance},f} = x_f(t_{\textnormal{touchdown},f}) - x_f(t_{\textnormal{liftoff},f})
\end{equation}
and for the slow limb:
\begin{equation} 
s_{\textnormal{stance},s} = x_s(t_{\textnormal{touchdown},s}) - x_s(t_{\textnormal{liftoff},s})
\end{equation}
The step length is the interlimb distance between the toes, at touchdown of one of the limbs:
\begin{equation}
s_{\textnormal{step},f} = x_f(t_{\textnormal{touchdown},f}) - x_s(t_{\textnormal{touchdown},f})
\end{equation}
\begin{equation}
s_{\textnormal{step},s} = x_s(t_{\textnormal{touchdown},s}) - x_f(t_{\textnormal{touchdown},s}).
\end{equation}

Similarly, the step time is an interlimb parameter:
\begin{equation}
t_{\textnormal{step},s} = t_{\textnormal{touchdown},s} - t_{\textnormal{touchdown},f}
\end{equation}
\begin{equation}
t_{\textnormal{step},f} = t_{\textnormal{touchdown},f} - t_{\textnormal{touchdown},s},
\end{equation}
and the stance time is an intralimb parameter:
\begin{equation}
t_{\textnormal{stance},f} = t_{\textnormal{liftoff},f} - t_{\textnormal{touchdown},f}
\end{equation}
\begin{equation}
t_{\textnormal{stance},s} = t_{\textnormal{liftoff},s} - t_{\textnormal{touchdown},s}.
\end{equation}

Finally, we define the double support time, as the time when both limbs have ground contact. We define the double support time for both limbs, as shown by the grayed out area in Figure \ref{fig:gait_temporal_parameters}:
\begin{equation}
DS_f = t_{\textnormal{stance},f} - t_{\textnormal{step},s} \; , \hspace{0.3cm} DS_s = t_{\textnormal{stance},s} - t_{\textnormal{step},f}.
\end{equation}

\section{Derivation of the learning rule}\label{app:learning_rule}

If the loss, or evaluation function, is defined as:
\begin{equation}
    V = \frac{1}{2}e_t^2.
\end{equation}
Then an error-based update rule can be derived for the cerebellum:
\begin{equation}
    y_\textnormal{cerebellum}^{(i+1)} = y_\textnormal{cerebellum}^{(i)} - \alpha \frac{\partial V}{\partial y_\textnormal{cerebellum}}
\end{equation}
If we assume the step time to be a linear function of the phase duration, with a positive correlation constant $k_{corr}$:
\begin{equation}
    t_{\textnormal{step}} = t_{\textnormal{step}}^0 + k_{corr} (\Delta\phi_F - \Delta\phi_F^0),
\end{equation}
where $t_{\textnormal{step}}^0$ is the base step time given when $\Delta\phi_F = \Delta\phi_F^0$, then the gradient will be:
\begin{equation}
\begin{aligned}
    \frac{\partial V}{\partial y_\textnormal{cerebellum}} &= \frac{\partial V}{\partial e_t}\frac{\partial e_t}{\partial t_{\textnormal{step}}}\frac{\partial t_{\textnormal{step}}}{\partial \Delta\phi_F} \frac{\partial \Delta\phi_F}{\partial y_\textnormal{cerebellum}} \\
    &= e_t\cdot \frac{\partial e_t}{\partial t_{\textnormal{step}}} \cdot k_{corr} \cdot 1
\end{aligned}
\end{equation}
The remaining derivative is different for the slow and the fast limb. For the slow limb:
\begin{equation}
    \frac{\partial e_t}{\partial t_{\textnormal{step},s}} = \frac{\partial e_t}{\partial DS_f} \frac{\partial DS_f}{\partial t_{\textnormal{step},s}} = (-1)\cdot (-1) = 1,
\end{equation}
and for the fast limb:
\begin{equation}
    \frac{\partial e_t}{\partial t_{\textnormal{step},f}} = \frac{\partial e_t}{\partial DS_s} \frac{\partial DS_s}{\partial t_{\textnormal{step},f}} = 1\cdot (-1) = -1.
\end{equation}

The final cerebellar adaptation rules are then
\begin{equation}
    y_\textnormal{cerebellum, s}^{(i+1)} = y_\textnormal{cerebellum}^{(i)} - \alpha^* e_t,
\end{equation}
and
\begin{equation}
    y_\textnormal{cerebellum, f}^{(i+1)} = y_\textnormal{cerebellum}^{(i)} + \alpha^* e_t,
\end{equation}
where $\alpha^* = \alpha k_{corr}$ is the learning rate. This is the only parameter that has to be chosen for the algorithm.

\end{document}